\documentstyle[12pt,aps,pra,preprint]{revtex}

\draft
\tighten

\def\be{\begin{equation}}
\def\ee{\end{equation}}

\begin{document}

\title{QUANTUM SCATTERING IN THE STRIP:  \\
FROM BALLISTIC TO LOCALIZED REGIMES}

\vskip 0.5truecm

\author{ Robert G{\c e}barowski$^{1}$\footnote{Present address:
Department of Applied Mathematics and Theoretical Physics,
The Queen's University of Belfast, Belfast BT7 1NN, Northern Ireland}, 
\ Petr \v{S}eba$^{2,3}$,  Karol
\.Zyczkowski$^4$\footnote{Fulbright Fellow. Permanent
address:
Instytut Fizyki im. Mariana Smoluchowskiego, Uniwersytet
Jagiello\'nski,
  ul. Reymonta 4, 30-059 Krak{\'o}w, Poland},
 and  Jakub Zakrzewski$^5$}

\address{
$^1$Instytut Fizyki, Politechnika Krakowska im. Tadeusza
Ko\'sciuszki, \\
ul. Podchor\c{a}\.zych 1, 30--084 Krak\'ow, Poland.}
\address{
$^2$ Nuclear Physics Institute, Czech Academy of Sciences,
\\
250 68 {\v R}e{\v z} near Prague, Czech Republic }
\address{
$^3$ Pedagogical University, Hradec Kr\'alov\'e, Czech
Republic }
\address{
$^4$ Institute for Plasma Research, University of Maryland, \\
  College Park, MD 20742, USA}
\address{
$^5$Instytut Fizyki im. Mariana Smoluchowskiego, Uniwersytet
Jagiello\'nski,
\\ ul. Reymonta 4,  30-059 Krak\'ow, Poland}
\date{\today}
\maketitle

\begin{abstract}
Quantum scattering is studied in a system
consisting
of randomly distributed point scatterers in the strip.
The model is continuous yet exactly solvable.
Varying the number of scatterers (the sample length) we investigate
a transition between the ballistic and the localized
regimes. By considering the cylinder geometry and introducing
 the magnetic flux we are able to
study time reversal symmetry breaking in  the system.
Both macroscopic (conductance)
and microscopic  (eigenphases distribution, statistics
of $S$ matrix elements) characteristics of the system
are examined.
\end{abstract}


\vskip 2truecm

\newpage
\section{Introduction }

A rapid technological progress in producing small
semiconductor and metallic samples, which have the size of the electron
phase coherence length, has stimulated extensive investigations
of transport through disordered mesoscopic systems \cite{ALW91,AMPZ95}.
Disorder in mesoscopic nanostructures, like for example 
quantum wires, can be induced by  adding  some impurities,
 which serve as electron scatterers
and hence modify the electron transport properties.
The sample length $L$ and the elastic mean free path, 
$l_e$ are important parameters 
for classification of various
possible regimes of scattering. In a sample, which size $L$ 
is of the order of $l_e$ (the ballistic regime),
the  traveling electron encounters only a few
scattering events on its path. However, for samples much longer 
than the mean free path, the multiple scattering
becomes increasingly important. 
In addition to that, the number of degrees of freedom
in the system and the energy of the incoming
electron also determine the transmission properties.

Anderson localization is probably the most important phenomenon relevant
 for such studies \cite{Anderson58}. 
 Its generality became fully appreciated
after the formulation of the scaling theory \cite{abr79}.
In particular, this theory predicts that extended states (which may
lead to the metallic, quantum diffusive regime) are possible
only for three-dimensional (3D) and weakly disordered systems, 
while in one (1D) or two--dimensional (2D) systems the states are
exponentially localized (in the absence of external fields).
Special topical meetings devoted to the Anderson localization
\cite{Ando88} have covered both theoretical aspects and
experimental manifestations of that phenomenon.

A considerable progress has been recently made in  
understanding transport properties in disordered samples
on grounds of the Random Matrix Theory (RMT) \cite{Mehta}
(for a recent review see \cite{Bena97}).
Moreover, the RMT approach has been successful in explaining statistical
properties of fluctuations occurring in the regime of parameters
where the corresponding classical
scattering exhibits chaos in the interaction region (the so called
chaotic scattering) \cite{BS88,Smil90,BM94,JPB94}.
More general  ensembles of random matrices
were used to describe quantum systems 
in the localized regime \cite{MI95}.
In such systems
classically chaotic motion
corresponds to 
 the disorder in mesoscopic nanostructures.

Not only were the transport phenomena
related to electrons under experimental
and theoretical scrutiny. A great deal of
effort by theorists and experimentalists has been
devoted to the electromagnetic wave transport
through microwave cavities and guides
\cite{BS88,BM94,Stoeckman,FS97}. Let us mention that
the Anderson localization
has been also observed for light scattered in a random
medium \cite{wblr97}. 

So far, numerical studies of
 a transition from the
ballistic to the localized regime have been mainly restricted
to tight-binding models. As typical examples one may consider
 \cite{BG92,BG93} for the 1D case,
\cite{JP95} for the 2D  model or  \cite{Slevin93} for 3D calculations. 
Nevertheless, such models may be regarded as discrete approximations to
a continuous scattering problem. While fully appreciating all the
results obtained within these approximate models, it is
desirable to compare them with those coming from exactly solvable systems
(if such are available). In this paper we shall consider
such an exactly solvable system, namely a two dimensional
strip with $N$ randomly distributed point scatterers. The
system has been  recently introduced in \cite{EGST96}, where it has
been shown that, the $S$--matrix eigenphases
obey the Wigner near neighbour statistics for a particular choice of parameters. 
The aim of present paper is to show a  more detailed study 
of that model, which reveals a smooth transition from the ballistic to
localized regime as a number of scatterers (or the sample length)
is varied. We investigate both the
macroscopic (conductance $G$) and
the microscopic properties of the system (statistical properties
of the $S$--matrix).

It is known for a long time that in the 2D case a proper quantum diffusive
regime does not exist typically \cite{abr79}.
However we show that in the
transition regime, which is beyond   
 the ballistic regime and yet  far from the localized one,
the system exhibits similar properties to those expected when
the multiple scattering occurs. Thus we 
are able to observe some resemblance to
the behaviour which is typical for the quantum diffusive regime.
Let us mention here that this finding is in full
agreement with earlier studies \cite{BG92,BG93,JP95}.

The 2D strip model is relevant in description of the
disordered transport when the time reversal symmetry is conserved. 
However this symmetry can be easily destroyed
by introducing a magnetic field into the system.
In order to demonstrate the capability of our model,
we show also that
by a change of the geometry and considering a
cylinder with the axial magnetic field, it is possible to 
extend our solvable approach so as to be able to investigate
the influence of the magnetic flux inside the cylinder
on the scattering occurring on its surface.

The paper is organized as follows.
Section II gives an overview of the both versions of the model:
with and without the magnetic field.  In this section we define 
 important
physical parameters characterizing the scattering system.
The conductance within the model is studied in Section III,
whereas the $S$--matrix properties are presented in Section IV.
Finally, the last section brings the summary and
conclusions.

\section{The scattering model}

A disordered mesoscopic sample is  modeled by a 2D
hard--wall strip with a finite number $N$ of point
scatterers \cite{EGST96}.
The geometry of the strip (the scattering region) is described by its 
length $L$, which can be varied, and its width which is
 set at the value $W=\pi$
(results for an arbitrary $W$ may be obtained by a simple rescaling
\cite{EGST96}).
 The point--like scatterers are randomly and uniformly
distributed inside the strip with positions
 ${\bf x}_j = (x_j, y_j)\in
 (-L/2, L/2) \times (0, \pi)$, for $j=1,...,N$
 (as shown in  Fig.~\ref{schemat}(a)).

It is assumed that the scattering in the strip is elastic and
the wavefunction vanishes on the horizontal strip boundaries
(hard walls) so that
\be
 \psi(x,0) = \psi(x,W) = 0, \qquad {\rm for \ \ all} \ \ x.
\label{dirichlet}
\ee
Appropriate boundary conditions \cite{EGST96} are also set on
small circles of radius $a_j$ surrounding the j-th point scatterer.
Everywhere except the inside of the circles the Hamiltonian of
the system corresponds to the free propagation, $H_0=-\Delta$
(in the units  $\hbar/(2m)=1$). Together with the boundary
conditions mentioned above this procedure allows for a
rigorous construction of a self-adjoint extension 
of the Hamiltonian in the presence of singular point perturbers.
We refer the mathematically oriented reader to \cite{EGST96}
for details of this particular model, while the extensive discussion
of mathematical techniques used to solve singular point interaction
problems may be found in \cite{AGHH}. For the purpose of the present study,
 it suffices to say  that the 
real parameters $\alpha_j$ characterizing the
self adoint extension are related to $a_j$ by $\alpha_j = -\ln(a_j)/2\pi$
\cite{EXSE}.
 In the following we shall take  
these perturbers to be identical by setting $\alpha_j=\alpha$ for
all $j$. A large positive value of $\alpha$ corresponds to a weak perturbation,
and in particular the perturbation can be removed by taking
 a limit $\alpha \rightarrow \infty$   
 so that the self adjoint extension is simply equal to $H_0$.

 The number of channels $M$
in which the electron may enter the system from either side
of the strip is equal to the integer part of the length of the
total wave vector $\vec{k}$ of the incoming electron.
Let $a^l_{in}$ and $a^r_{in}$ denote the $M$--component vectors
representing the waves incoming from the left and the right side
of the strip, respectively
(depicted in Fig.~\ref{schemat}(a) as the two
arrows pointing towards the strip from left and right).
The outgoing waves from the left and right side of the strip
(shown in Fig.~\ref{schemat}(a) by the two arrows
 pointing outwards) are described by
$a^l_{out}$ and $a^r_{out}$, respectively.
The scattering process can be described by a $2M \times 2M$
unitary matrix $S$. The $S$--matrix
 relates amplitudes of the incoming waves with those which represent
 outgoing ones so that
 $\{ a^l_{\rm{out}}, a^r_{\rm{out}} \}=S
\{ a^l_{\rm{in}}, a^r_{\rm{in}} \}$
and  therefore it has the following block structure
\be
S = \left (
     \begin{array}{cc}
       r   &  t \\
        t' &  r'
     \end{array}
\right ),
\label{Smat}
\ee
where $r$ and $r'$ are reflection matrices whereas
$t$, $t'$
are transmission sub--blocks, each of them having size $M \times M$.

The advantage of the model with singular point--like perturbers is
that explicit  formulae for the
 $S$--matrix elements are available \cite{EGST96}
\be
r_{nm} (E) = \frac{i}{\sqrt{2 \pi}} \ \sum_{j,k=1}^{N}
[\Lambda(E)^{-1}]_{jk}
\ \frac{\sin(m y_j) \sin(n y_k)}{\sqrt{ k_m(E) k_n(E) }} \
\exp[ i(k_m x_j + k_n x_k)]
\label{rnm}
\ee
and
\be
t_{nm} (E) = \delta_{n,m} \ + \ \frac{i}{\sqrt{2 \pi}} \
\sum_{j,k=1}^{N} [\Lambda(E)^{-1}]_{jk}
\ \frac{\sin(m y_j) \sin(n y_k)}{\sqrt{ k_m(E) k_n(E) }} \
\exp[ - i(k_m x_j - k_n x_k)] ,
\label{tnm}
\ee
where $E$ is the energy of incident electrons. The  $N
\times N$
matrix $\Lambda(E)$
is given by its elements
\be
\Lambda_{jj} (E)  = \alpha + \frac{1}{\pi} \
\sum_{n=1}^{\infty} \left [
 \frac{1}{2n} - \frac{i\sin^2(n y_j)}{k_n(E)} \right ],
\label{lambdadiag}
\ee
and
\be
\Lambda_{jm} (E)  =  - \frac{i}{\pi} \ \sum_{n=1}^{\infty}
\frac{\exp( i k_n(E) |x_j - x_m|)}{k_n(E)} \sin(n y_j)
\sin(n y_m) , \qquad j \neq m.
\label{lambdaoff}
\ee
The longitudinal momentum $k_n$ satisfies  the relation
\be
k^2 = k_n^2 + n^2,
\label{klong}
\ee
hence for $n > k$, it becomes imaginary $(k_n\sim i n)$,
what ensures the convergence of the series~(\ref{lambdadiag}).

Conductance in the strip can be calculated using the
famous Landauer formula
\be
G  = G_0 \sum_{n,m=1}^M {\rm Tr} \{tt^{\dagger}\},
\label{landauer2}
\ee
where $G_0= e^2/h$ (we omit the spin degeneracy factor).
Since the matrix $S$ is  unitary,
Tr$\{tt^{\dagger}\}=$Tr$\{t't'^{\dagger}\}=M-$Tr$\{rr^{\dagger}\}$, so
the transmission (reflection) coefficients for electrons
entering the strip from both sides are equal.
In the following we work in dimensionless units and omit the
proportionality factor $G_0$.

In order to break the time reversal symmetry we
introduce a magnetic field into the model \cite{EGST96}.
It is convenient to modify the strip with scatterers
 into a cylinder of length $L$ and perimeter $2 \pi$,
  as shown in Fig. \ref{schemat}(b).
 Periodic boundary conditions read now

 \begin{eqnarray}
  \psi(x,0) & = & \exp \left (i \frac{\Phi_B}{2\pi}\right )
\ \psi(x,2\pi) \cr
  \frac{\partial }{\partial y} \psi(x, 0) & = &
        \exp \left ( i \frac{\Phi_B}{2\pi} \right )
 \frac{\partial}{\partial y} \psi(x,2\pi),
 \label{periodicbound}
 \end{eqnarray}
where $\Phi_B$ is the magnetic flux
inside the cylinder.
Since the cylinder can be mapped onto the rectangle $\Omega  =
 [0, L] \times [0, 2\pi) $, the solutions obtained
in the case of the strip can be modified accordingly
--- for more details see \cite{EGST96}.
The longitudinal momentum depends in this case on the
magnetic flux, so Eq.(\ref{klong}) has to be replaced by
the following 
\be
k_n(E) = \sqrt{E - (n + \Phi_B/(2\pi))^2},
 \qquad n=0, \pm 1,\ldots, \pm M.
\label{kmb}
\ee

Let us point out that both versions of the model are continuous and
exactly solvable in the sense that the scattering matrix
elements can be written analytically (albeit in terms of
infinite series).
We believe therefore that this makes the model extremely interesting 
and useful for investigating the statistical
properties of $S$--matrices and for analysis of
conductance in disordered mesoscopic media.

In order to get a proper statistical sample of data we average
the studied properties over
several realizations of random positions
of $N$ scattering points  for both versions of the
model (the strip and the cylinder with the magnetic flux).
 We choose
the local density of the scattering points, $\rho$, to be constant by
setting $L=N$ (this choice yields the 
scatterers density $\rho=1/\pi$ for the strip and twice as smaller
value for the cylinder model). Hence throughout this paper we shall use
the sample length $L$ to parametrize the model.
We consider the perturbers to be strong and take $\alpha=0$
(recall that $\alpha\rightarrow\infty$ corresponds to a vanishing
perturbation).

As it has been pointed out in the Introduction,
the elastic mean
free path $l_e$ is yet another important parameter which
determines the regime of the scattering behaviour. It is therefore  
crucial to establish its value for the discussed model.
 The total cross--section $\sigma$ for the
scattering on a single point--like impurity in 2D may be
obtained from an appropriate formula valid for the scattering in the
plane \cite{AGHH} and for $\alpha=0$ it reads
\be
\sigma=\frac{\pi^2}{k}\frac{1}{[\gamma+\ln(k/2)]^2+\pi^2/4} ,
\label{cross}
\ee
where $\gamma\approx0.577... \ $ is the Euler constant. 
The mean free path itself can be expressed in terms of
the total cross section and the density of point--like impurities
$l_e=1/\rho\sigma$.
Note that, because the density $\rho$ is constant
as a result of the choice $N=L$, the mean free path
has a fixed value for a given incident energy (or the value $k$ of
the wave vector). This allows for approaching
various regimes of the scattering simply by changing
the length of the strip together with  the number 
of the randomly distributed impurities.

\section{Conductance of the strip}

To achieve the ballistic (or quasi--ballistic) regime, the sample length $L$
(equal, in our model to the number of point like scatterers $N$)
should be of the order of the elastic mean free path $l_e$. The
latter strongly depends on the
wavevector $k$ (the integer part of which gives the number of
open channels $M$) --- compare Eq. (\ref{cross}).
 In most of the discussed numerical examples we shall
take $k=5.5708$, which corresponds to $M=5$. This yields $l_e\approx 8.9$
for the strip model.

Once we have that $N \gg l_e$, the multiple scattering in the sample 
becomes possible.
Incoming waves are strongly distorted in this regime. Since quantum
diffusion does not appear in 2D \cite{abr79} we expect a smooth
transition from the ballistic to the localized regime for
a sufficiently large $L=N$.

The $\Lambda$ matrix elements are
given by infinite sums,
compare Eqs.~(\ref{lambdadiag}-\ref{lambdaoff}).
For  realistic computation times (all calculations have been
performed on a personal computer) we have to restrict the analysis
 to moderate $L$ values.  For that reason we have chosen $M=5$ in most
of our numerical studies.

Fig.~\ref{condav}(a) shows on the logarithmic scale
the dependence of the mean conductance 
$\langle G \rangle$ 
on the length of  the strip $L$. Observe that  for $L\ge 100$,
 the conductance decays
exponentially with $L$ as expected for sufficiently long samples
in the localization regime.
A fit to  the exponential decay 
of the mean conductance, $\langle G \rangle \propto \exp(-L/\xi)$
 (shown in the figure with a broken line), yields an estimate for
the localization length
$\xi=100\pm 3$.  The mean conductance for shorter
samples is presented in Fig.~\ref{condav}(b) in the double logarithmic scale.
Note that for  values $L\in(10,100)$,
a straight line well approximates the data, thus yielding
the power--low dependence on the sample size $L$,
$\langle G \rangle\propto L^{-c}$ with $c=0.90 \pm 0.05$. This is not, however,
an indication of the quantum diffusive regime 
(where according to the theory $\langle G \rangle$ should be
proportional to $L^{-1}$), but it rather stems from the fact that the sample is
shorter than the localization length $\xi$. Thus the localization
(depending on the quantum interference occurring on the length scale of $\xi$)
cannot fully set in. Therefore,
 what we really observe here is a transition from the
ballistic regime (dominated by direct processes) to a fully localized
situation.

The introduction of point perturbers into
a classically integrable system leads to a seemingly  
``quantum chaotic'' behaviour for a finite value of $\hbar$ --
the so called wave chaos \cite{seba90,Seba97a}. Hence, it is interesting
to compare the behaviour 
of our model with point scatterers with that predicted for chaotic
scattering models. For the sake of such analysis, it
has been conjectured that the Random Matrix Theory \cite{Mehta}
 approach correctly captures the statistical
properties of the $S$--matrix \cite{BS88,BM94}
(for the equilibrated component -- using
the nuclear physics language \cite{lw92}, or after unfolding the $S$
matrix -- see next Section -- using the language typical for RMT
applications to bound systems).
On the ground of this conjecture
statistical properties of the  unitary  matrix $S$
may be represented by random matrices of Dyson circular 
ensembles \cite{Dy62}.

Leaving a more detailed microscopic comparison for the next Section,
let us consider here
 the mean conductance and its variance. For the $S$ matrix pertaining to
 the Circular Orthogonal Ensemble (COE), which is relevant
 for the strip model preserving the time--reversal symmetry, 
 the RMT yields a value for the mean conductance
(including the weak localization correction) 
and its variance \cite{Bena97,BM94} which may be expressed in terms of the number
of channels $M$ as follows
\be
\langle G \rangle_{\rm R} = \frac{M}{2} -
\frac{M}{4 M + 2},
\label{gavrmt}
\ee

\be
{\rm Var}_{\rm R} (G) =
\frac{M (M + 1)^2}{(2 M + 1)^2 (2 M +3)}.
\label{gvarmt}
\ee
 The above formulae yield
$\langle G \rangle_{\rm R} =2.27 $ and  ${\rm Var}_{\rm R} (G)=0.1144$,
when the number of channels $M=5$ is taken.

Observe that while the mean conductance rapidly passes through
the RMT value (depicted with a horizontal solid line in
both panels of Fig.~\ref{condav}), the behaviour of the variance
is markedly different (compare Fig.~\ref{condav}(b) and Fig.~\ref{condvar} 
-- in the latter the value of ${\rm Var}_{\rm R} (G)$ is depicted by
 a dashed horizontal line).
The variance rapidly increases when $L$ exceeds the value corresponding
to the mean free path $l_e$, then saturates
in the vicinity of the RMT value and afterwards 
again rapidly falls down for $L>\xi$.
Thus in the ``transition'' region defined by the interval  $l_e<L<\xi$,
a rapid, almost linear change on the logarithmic scale in $\langle G \rangle$ is
accompanied by a relatively weak dependence of the ${\rm Var}(G)$ on
$L$ --- a phenomenon closely resembling universal conductance fluctuations
in metallic samples.

In our case the direct processes strongly
influence $\langle G \rangle$. Intuitively, in the transition region they should
affect the variance to a lesser extent. In this region the localization
does not set fully yet (since the sample is too short) and we observe
the multiple scattering (typical for a chaotic process) which manifests
itself in the value of the variance being close to the RMT prediction.
We mention, however, that for the problem studied we are not
in the semiclassical regime since the number of open channels
is too small as well as the singular perturbers do not have a 
 proper classical limit
(that is in such a limit, their perturbing effect disappears completely).

In the localization regime, the conductance becomes vanishingly
small, it is therefore much more informative to study not the
conductance itself but its logarithm. 
In Fig.~\ref{fig33} both the mean logarithmic
conductance $\langle -\ln G \rangle$ (asterisks)
and its variance (circles) are presented.
For a strip length taken from the interval $L \in (25, 100)$
 (the transition region)
both the mean and the variance of the logarithmic
conductance take approximately the same values. On the other hand,
for $L> 200$ there is a clear growth of the variance
${\rm Var}_{\rm ln}=\langle [ \ln G - \langle \ln G \rangle ]^2\rangle$
 which satisfies in this range of $L$ values 
the relation
\be
  {\rm Var}_{\rm ln}  = 2 \langle - \ln G \rangle
\label{varlok}
\ee
predicted for localized regime \cite{Bena97,Slevin93,DP,italo} as
shown in the inset with a solid line.
In fact, the linear regression fit for the numerical data
yields the slightly smaller value $1.90 \pm 0.09$ for the slope.

Not only the mean conductance and its variance but also
the distribution of the conductance may be studied.
According to the predictions of the RMT
\cite{Bena97,Slevin93}, in the quantum diffusive regime
the distribution of the conductance, $P(G)$,
is a Gaussian for a large number of channels $M \gg 1$.
Nevertheless, some results for the transport through
 chaotic cavities \cite{BM94}
indicate that already for $M=3$ the conductance distribution
appears to be approximately a Gaussian.
We observe a similar behaviour in the transition region
(between the ballistic and localized regimes) discussed above.
Exemplary data for such a case are displayed in Fig.~\ref{fig4}(a),
where the conductance distribution is shown for $L=20$
(it has been obtained from $10^4$ configurations of random scatterers).
In the figure a thin solid line shows a Gaussian
distribution with the mean and its variance calculated from
that statistical representation.
Thus not only the
variance but the whole distribution of the conductance in the
transition region behaves in the way associated typically with
the metallic regime.

The distribution $P(G)$ changes with
an increase of the number of the scatterers (and the length of the strip)
and once it reaches the localized regime, it can be
approximated by a log--normal distribution  \cite{Bena97,Slevin93,Imry86}
\be
P( \ln G) = \frac{1}{\sqrt{2 \pi \sigma^2_{\ln} }}
\exp\bigl[ - { (\ln G - \langle \ln G \rangle)^2 \over
   2 \sigma^2_{\ln} } \bigr]
\label{pglok}
\ee
with the mean and variance related to each other by Eq. (\ref{varlok}).
The distribution (\ref{pglok}) of the logarithmic
conductance is due to the nature of the localization, where
the total wavefunction can be super--imposed of partial
wavefunctions, which have very small overlaps.
Thus, the total transmission
probability is given by a product of component transmissions,
which explains the origin of a Gaussian distribution of
logarithms of the conductance.

In Fig.~\ref{fig4} -- panels (b) and (c) -- we present
the distribution of the logarithmic conductance for $L=200$ and $L=400$,
respectively. Let us mention that the log--normal distribution
(\ref{pglok}) (depicted in both panels by a thin solid line)
well describes the conductance fluctuations only in the latter case.
Although in Fig.~\ref{condav} we have seen clearly the indications
of the localization occurring for $L>100$, 
it fully sets in for much longer samples for which 
additionally the logarithmic conductance obeys closely
 the log--normal distribution.

Let us partially summarize the results obtained directly from
the studies of measurable quantity -- the conductance of the strip.
For short samples, $L\approx l_e$, the scattering is dominated
by direct processes --- the ballistic scattering occurs. For
long samples the mean conductance decreases exponentially with $L$
yielding the localization length $\xi\approx 100$ (for $M=5$ case
studied). The relatively broad transition region $l_e<L<\xi$
exhibits a behaviour reminiscent of the quantum diffusion in 3D
metallic samples: an approximately linear
 decrease of the mean conductance $\langle G \rangle$ with 
 the inverse sample size $L^{-1}$ (the Ohm law, a similar
behaviour for a tight binding 1D model has been reported in \cite{BG92}),
the variance is roughly independent of the mean conductance (an
analog of the universal conductance fluctuations), 
and finally the conductance  has a Gaussian
distribution. It is interesting to see
whether the microscopic properties ($S$--matrix elements, its
eigenphases) likewise support that description of the scattering regimes.

\section{Microscopic properties}

The most well known tool in studies of statistical properties
of a sample of levels is the nearest neighbour spacing distribution
\cite{Mehta,Haake}. For an ensemble of unitary $S$ matrices,
  the corresponding
measure is the distribution of eigenphase spacings (i.e. writing
the eigenvalues as $\exp(i\theta_j)$ we consider the distances
$\theta_{j+1}-\theta_j$ for $j=1,\ldots,(2M-1)$, for a $2M\times 2M$
scattering matrix). For the COE a good approximation for the
eigenphases spacing distribution is the Wigner surmise
\cite{Haake}. Such a distribution is expected also for the
chaotic scattering occurring in the ballistic regime \cite{BS88,BM94},
provided that the direct processes are eliminated (i.e.
the processes which contribute to  a non--vanishing
average over the ensemble, $\langle S \rangle \ne 0$).
In the case of $\langle S \rangle=0$ one expects an  uniform distribution of
eigenphases over the interval $[0, 2\pi)$.

When the direct processes are important one should first
eliminate their contribution from the $S$ matrix (i.e. ``unfold'' this matrix)
by considering
$S_{\rm fl}= S-\langle S \rangle$, where $\langle S \rangle$ is
the matrix obtained by an arithmetic mean of $S$ matrix elements over
different realizations of the disorder.
Several procedures for ``unfolding'' the $S$ matrix have been
proposed
\cite{BG93,lw92,fm85} -- we have followed the prescription of Friedman
and Mello \cite{fm85}.
 Let us mention that in \cite{JP95}
another unfolding procedure, borrowed from studies of bound systems
\cite{Bohigas} has been used.

Let us consider first
 an extremely short sample of the length $L=5$.
The resulting spacing distribution $P(s)$ after the unfolding procedure 
is shown
in Fig.~\ref{spac1} (a heavy line). It is well approximated 
neither by the Poisson (a dashed line) nor the Wigner distribution
(a thin solid line), though
the latter far better fits to the data than the former
one especially for small spacing values $s$.
For $L<l_e$,  at
most one scattering event perturbs the incoming wavefunction,
so there is no reason to expect a good agreement with RMT predictions.

As soon as we enter the transition region ($l_e<L<\xi$, see
the previous section), the direct scattering ceases to be the
dominant process, and the multiple scattering comes into play.
The  initial
wavefunction becomes strongly distorted while passing through
the sample and indeed the spacing distribution is much better
approximated by the Wigner distribution (compare Fig.~\ref{spac2}
displaying data for $L=40$).
This is consistent with the discussion presented the previous section
(a Gaussian distribution of the conductance, an approximate independence of
its variance on $L$).

As mentioned above, in the transition region, where
the  multiple
scattering is important, a coherent backscattering should appear
for the time reversal system due to interference between multiple
scattering paths passing in the opposite directions. This effect
is a manifestation of the ``weak localization'' and has been
observed in earlier studies, both for electron \cite{Ando88,bergmann82} and
light waves \cite{wblr97}. In the 3D case it appears for energies above
the transition to the localization (when $kl_e>1$).

A similar effect can be seen when $S$ matrices are treated within the
RMT approach.
The mean squared modulus of matrix elements averaged over the COE/CUE
ensembles (the latter being the Circular Unitary Ensemble)
reads
(eg. \cite{Bena97})
\be
\langle |S_{mn}|^2 \rangle_{\beta} = \frac{1 - (1-2/\beta) 
\delta_{mn}}{2\ M -1 +2/\beta} ,
\label{scoue}
\ee
with $\beta=1$ and $2$ for the COE and CUE, respectively.
The presence of an enhanced diagonal in this
distribution for $m=n$
is a distinct footprint of the time reversal symmetry (see also
\cite{MAS88,MS91}).

In order to test whether the weak localization effect appears in the
transition region we turn to the cylinder model since it
allows us to perform breaking of the time-reversal symmetry by introducing
 the magnetic flux (see Section II).
In Fig.~\ref{fig8mss} the scattering probabilities $|S_{mn}|^2$,
averaged over an ensemble of $10^{3}$  $S$--matrices
(each of them corresponding to a particular configuration of random point
scatterers)
 are presented for the case
corresponding to the fixed momentum $k=10.5708$
and the cylinder length $L=200$. Since the density of scatterers is
$\rho=1/2\pi$, using (\ref{cross}) we obtain $l_e\approx 50$.
Indeed, the enhancement of the backscattering
to the same channel is clearly visible in the case of null magnetic flux
(see panel (a) -- the anti--diagonal of the reflection sub--blocks
rather than the diagonal elements are twice as larger than the
off--diagonal elements due to the choice of the channel basis
representation \cite{EGST96} suitable for the cylinder version
of the model).
The selective enhancement of the reflection is destroyed
as the axial magnetic field is turned on
as demonstrated in panel (b), showing results for the magnetic flux
$\Phi_B/(2\pi) = 0.1$.

It is worth emphasizing that in Fig.~\ref{fig8mss}  there is also
visible a clear enhancement on the diagonal of the transmission matrix
which is an indication of the presence of direct processes
(the data presented are not unfolded). Also the off--diagonal
elements of reflection matrices are several times bigger than
the transmission matrices elements.

Let us consider now the limiting case of an extreme  localization.
The incoming wave is fully reflected, that is
$t$,$t'$ $\approx 0$
and the reflection sub--blocks $r$,$r'$ are unitary (this
property of $S$ matrix was discussed earlier in \cite{BG93,JP95}).
The reflection is a result of complicated scattering events, hence
both matrices $r$,$r'$ may be considered as those pertaining  to
the COE (or CUE for the cylinder model with the magnetic field).
Actually, as discussed in detail in \cite{JP95}, such a
situation occurs for quasi-1D samples only, the transverse
dimensions of the sample lead to deviations from RMT predictions.
Recall, however, that we consider the strip of width $W=\pi$
and the length $L \gg W$ so indeed we may consider
this to be a quasi--1D situation.

The scattering matrix $S$, Eq.~(\ref{Smat}),
 in the fully localized case,
consists thus of two independent diagonal blocks, each pertaining to an
appropriate circular ensemble.
Therefore spectral properties of the entire $S$ matrix  correspond 
to a superposition of two independent COE (or CUE) spectra.
The level spacing
distribution for such a case does not reveal any usual effect of level
repulsion but it is rather described by the
Berry--Robnik distribution \cite{BR84}
for two disconnected chaotic regions of equal volume. While in
\cite{BR84} the time-reversal invariant situation is mainly
discussed, the same formalism is easily extended for unitary
ensembles. Approximating the spacing distribution for the COE 
 by the corresponding Wigner distribution (exact, strictly
speaking for Gaussian $2\times 2$ matrices only \cite{Haake})
one easily obtains
\begin{equation}
P_O(s)= 
\frac{\pi}{8} s \ \exp\bigl(- \frac{\pi}{16} s^2 \bigr) \ 
{\rm erfc} \bigl(\frac{\sqrt{\pi}}{4} s \bigr)
 \  + \ \frac{1}{2} \exp \bigl(- \frac{\pi}{8} s^2 \bigr)
\label{oe}
\end{equation}
for the approximate spacing distribution in this case.
The standard form,
 ${\rm erfc }(z) = \frac{2}{\sqrt{\pi}} \int_{z}^{\infty} \exp(-t^2) dt $,
 for the 
 error function is used above.

Before presenting the numerical data
let us point out that the choice of the initial basis leading to the
particular form
(\ref{Smat}) of the scattering matrix is somewhat arbitrary.
Taking a different order of components in the vector describing the
outgoing wave we may write
$\{ a^l_{\rm{out}}, a^r_{\rm{out}} \}=S'
\{ a^r_{\rm{in}}, a^l_{\rm{in}} \}$
where the scattering matrix $S'$ now has the form
\be
S' = \left (
     \begin{array}{cc}
       t   &  r \\
        r' &  t'
     \end{array}
\right ).
\label{Sma2}
\ee
Both  forms (\ref{Smat}) and
 (\ref{Sma2}) of the scattering matrix  describe the same physical phenomenon.
However, in the extreme localization regime
($t, t' \approx 0$)
the statistical properties of $S$ and $S'$ are not the same.
The matrix $S'$ consists of two off--diagonal unitary blocks
$r$ and $r'$. We show in the appendix
that the eigenvalues of such a matrix are given by
\be
{\rm eig} (S')=\{ {\rm eig} (\sqrt{ rr'} ), - {\rm
eig} ( \sqrt{ rr'} ) \}.
\label{eigs}
\ee
If unitary matrices $r$ and $r'$ pertain to
the CUE (and are distributed uniformly according to the Haar measure),
so does their product $rr'$. As a result of that, the spectrum
of $S'$ consists 
of two replicas of a CUE--like spectrum, rescaled by the
factor of two.

Let us assume that, for the time reversal version of the model,
$r$ and
$r'$ can be described by the COE. Using a concept of composed 
unitary ensembles it was
shown that in such a case
the product $rr'$ fulfills a weaker property. It displays the COE-like
spectrum, in spite of the fact,
that it does not pertain to the COE \cite{PZK97}.
Thus the local statistical properties of $S'$ (analyzed at the
scale of mean level spacings $d \ll M$)
 are the same as those of the corresponding circular
ensembles, in contrast to the properties of $S$.

The spectral properties of $S$ and $S'$ are different
also in a realistic case of a nonzero conductance.
To demonstrate that we
present in Fig.~\ref{spac3} numerical results obtained for
$L=800$, i.e.  far into the
localization regime.
 Panel (a) shows the spacing statistics
 $P(s)$ for the matrix $S$ whereas panel (b)
 displays that statistics for the $S'$--matrix.
 It is evident that the spacings distributions
for $S$ and $S'$ are indeed very much distinct,
the former
being quite reasonably approximated  by (\ref{oe}) while
the latter by the Wigner distribution (a thin solid line).
It has to be emphasized that this supports the
conjecture that two uncoupled
COE--like matrices may represent both reflection sub--matrices.

This finding is also consistent with tight--binding model 
calculations \cite{JP95},
where similar agreement has been found for quasi--1D situations.
It has been shown in that paper \cite{JP95} that when ``transverse'' dynamics
becomes important the number variance of $S$ matrix eigenphases,
$\Sigma^2(d)$,
exceeds that for two uncoupled COE's. As discussed in \cite{JP95}
only $0<d \ll M$ range is interesting since all $2M$ eigenphases must
fall in  $(0,2\pi)$ interval. Hence, for the discussed numerically case,
$M=5$, we are limited to low values of $d$ only. 
As shown in Fig.~\ref{sigma}, in this
interval we observe quite nice agreement between $\Sigma^2(d)$ for
reflection matrices of the strip model with $L=800$ (a heavy solid line)
 and predictions for the COE (a thin solid line).
However, for shorter samples, $L=200$ (a heavy dashed line),
where the localization is not fully set as discussed in the previous section,
the number variance displays significant deviations from the COE
behaviour for $d > 1.5$.
Due to a nonzero conductance 
the reflection matrices $r$ and $r'$ are not unitary,
but the statistics of the eigenphases can still be described
by random matrices of circular ensemble. This is no longer true for
a larger conductance (shorter length $L$ of the strip).

\section{Conclusions}

In this paper we have discussed quantum
scattering and the transition from the ballistic
to the localized regime  in the
explicitly solvable model of point scatterers in the strip.
The model reproduces the Gaussian
 distribution of the
conductance in the diffusive regime and the log--normal
distribution
when the length of the strip is large enough to
reach the localization regime. The number of point
scatterers
which is necessary for representing the localized regime
strongly increases with the momentum of the incident
electron (the number of open channels).

We find that the statistics of eigenphases of the $S$-matrix
is of the Wigner type not only, as expected, in
the diffusive regime, but provided a specific structure of the $S$--matrix
is taken, it may also reveal the distinctive level repulsion property
 in the localized regime.
This is in contrast to the case
of quantized  autonomous  chaotic systems
for which dynamical localization manifest itself by a Poissonian like
level statistics \cite{Haake}. Therefore, our results demonstrate 
 that the spectral properties of the scattering matrix cannot
be used as a sole criterion for the  localization.
On the other hand, the localization in the system manifests itself
in the statistical properties of the reflection  and transmission
submatrices of $S$.

The results obtained within present model are in full agreement with
similar earlier studies \cite{BG92,BG93,JP95}. However, while
those studies were based on  tight binding models (which necessarily
discretize continuum) the present model is continuous and yet
 explicitly solvable.

The model allows one to include also effects due to
 a  magnetic flux and thus to follow
a gradual breaking of the time
reversal symmetry.
It is also possible to introduce a geometric reflection symmetry
(for example by drawing randomly the position of $N/2$ scatterers
only and taking the
remaining half as their mirror image with respect to the symmetry
line) and analyze the symmetric case discussed earlier in \cite{GMMB96,Z97}.
In addition,  the model may be used to analize other statistical
properties such as the distribution of Wigner time-delays
\cite{SZZ96,FS97,GM98} or other parametric properties of the $S$ matrix.

Finally, we would like to mention that the results of the model discussed 
in the present paper, may be relevant also for experimental studies. 
These might include systems like quantum wires doped with a number
of impurity atoms, whose configurations can be randomize
by a thermal process \cite{Mailly92}. In such a system,
the conductance and its variance behaviour could be investigated
with an increasing number of impurity atoms (and the length of the sample).
Another possibility is to carry out measurements
in a microwave domain, investigating the propagation through 
 waveguides with antennas \cite{AHK96}, where 
 breaking of the time reversal symmetry
has been recently realized experimentally for microwave
billiards \cite{Stoff95}. 
In addition, an experimental study of microscopic properties 
of the scattering matrix also could be feasible, for example
by looking at the backscattering to the same channel
as compared to the reflection to any other channel,
in the presence and absence of the time reversal symmetry.

\section*{Acknowledgments}

One of us (RG) wishes to acknowledge financial
support through the fellowship awarded by the Foundation ``Nadace pro
podporu teoretick\'e fyziky'' from Slemeno near
Rychnov nad Kn{\v e}{\v z}nou in the Czech Republic
and the hospitality at Instytut Fizyki, Uniwersytet
Jagiello\'nski in Krak\'ow. Another of us  (K\.Z)  thanks  Ed Ott
for hospitality at the Institute for Plasma Research,
University of Maryland, where part of this work has been done
and acknowledges the Fulbright fellowship.
Financial support by
Komitet Bada{\'n} Naukowych and the grant GAAV CR 1048804 is also
gratefully acknowledged.

\newpage

\appendix
\section*{A}

Let us consider  the following algebraic lemma:

\vskip 0.5truecm

{\bf Lemma:}

Let $A$ and $B$ be $M \times M$ unitary matrices and let $C$
reads
\begin{equation}
C = \left(\matrix{0   & A   \cr B  & 0\cr}\right).
\label{A1}
\end{equation}
Then the spectrum of $C$
is given by

\begin{equation}
{\rm eig}(C)= \{ {\rm eig} (\sqrt{AB}), - {\rm
eig}(\sqrt{AB})\}.
\label{A2}
\end{equation}

\vskip 0.5truecm

{\bf Proof of the lemma:}

Let eigenvalues and eigenvectors of $\sqrt{AB}$ be denoted
by  $d$ and
$U$ respectively, so that
$d=U^{\dagger}\sqrt{AB} U$. Consider an $2M \times 2M$
matrix $X$
defined as

\begin{equation}
X = {1 \over \sqrt{2}}
\left(\matrix{U    & U\sqrt{AB} B^{\dagger} \cr
              -U    & U\sqrt{AB} B^{\dagger} \cr}\right).
\label{A3}
\end{equation}
It is easy to show that $X$ is unitary.
Direct computation allows us to verify that $X$ consists of 
eigenvectors of $C$, so that
\begin{equation}
X CX^{\dagger}  =
\left(\matrix{ d    &  0  \cr
               0    &  -d  \cr}\right),
\label{A4}
\end{equation}
which proves the lemma.

\hskip 4truecm $\Box$

\newpage


\begin{figure}
\caption{ Schematic diagrams  of both models
 analyzed throughout the paper. Case a)
represents the time reversal model consisting of
 $N$ point scatterers randomly distributed
over the strip of width $\pi$ and length $L$.
Panel b) shows a cylinder
of the perimeter $2\pi$  with $N$ point scatterers on its surface.
 The magnetic flux inside the cylinder breaks the time reversal
invariance.
}
\label{schemat}
\end{figure}
\begin{figure}
\caption{Panel (a): The mean conductance $\langle G \rangle$ as 
a function of the length of the strip $L$ 
(the number of point scatterers $N=L$) for $M=5$ open channels. 
The numerical data,
each representing an average over 400 realizations are represented
by filled circles. For a sufficiently
long sample the exponential decay of $\langle G \rangle$ with $L$
is observed. The dashed line yields the fit of the localization
length $\xi=100\pm 3$. Thin horizontal line shows, for a reference,
 $\langle G \rangle_{\rm R}
=2.27$ value assuming that $S$ pertains to the COE.
Panel (b) shows a part of the data in the transition between the
ballistic and the localized regimes. Observe that a power--law
dependence of $\langle G \rangle$ on $L$ quite well approximates
the behaviour of the mean conductance in this region. The horizontal
straight line represents the RMT value, the same as in panel (a).
}
\label{condav}
\end{figure}
\begin{figure}
\caption{The variance of the conductance distribution for the same
parameters
as in Fig.~ 
 \protect{\ref{condav}}.
 The dashed line corresponds to
the RMT prediction (see text). Note that in the transition region
${\rm Var}(G)$ remains close to the RMT value and is only weakly dependent
on $L$.}
\label{condvar}
\end{figure}

\begin{figure}
\caption{The mean logarithmic conductance $\langle \ln G \rangle$ (asterisks)
and its variance ${\rm Var}_{\rm ln}$ (open circles) as a function
of the strip length $L$ for $M=5$ open channels.
 The inset shows the dependence of the variance on the mean ---
 the theoretical prediction for the localized regime is depicted with the line.
 For more discussion see text.
}
\label{fig33}
\end{figure}

\begin{figure}
\caption{Conductance distributions, $P(G)$, obtained for $M=5$
and for the length of the sample (strip) $L$ equal to
a) $20$, b) $200$ and  c) $400$ ($N=L$ in each case as usual).
Data in panel (a) are drawn in the linear scale and are
 compared to the normal
distribution. Data in panels (b) and (c) obtained for the
localized regime are displayed as $P(\ln G)$, while the thin
solid line in both panels
 represents theoretically predicted log--norm distribution 
for the localized regime.
      }
\label{fig4}
\end{figure}

\begin{figure}
\caption{
 The spacing distribution  $P(s)$ obtained for unfolded eigenphases 
 (see text) of
the $S$ matrix in the situation
 when direct processes dominate
the scattering (the length of the strip $L=5$ is smaller than the
mean free path $l_e$, the number of open channels $M=5$).
The numerical data shown in the form of a
 histogram are compared  with the Wigner (thin solid
line) and the Poisson (dashed line) distribution.
 }
\label{spac1}
\end{figure}

\begin{figure}
\caption{Same as Fig.  
 ~\protect{\ref{spac1}}
 but data are for $L=40$, i.e.,
in the transition region between the ballistic and the localized regime.
 }
\label{spac2}
\end{figure}
\begin{figure}
\caption{
 Scattering probabilities
$|S_{mn}|^2$, averaged over $10^3$ configurations of random
points, for the cylinder model with $k=10.5708$ and $L=200$.
Panel (a) shows a  3D diagram for null magnetic flux,
$\Phi_B=0$,
while  panel (b) shows data for this system with
the  magnetic flux $\Phi_b/(2\pi) = 0.1$
inside the cylinder.  Note the difference in the diagonal
elements corresponding to the reflection blocks.
(The
anti--diagonals rather than the diagonals of sub--matrices
 $r$, $r'$ are enhanced
 due to a specific choice
of the channel ordering used in the case of the cylinder 
--- compare 
{\protect{\cite{EGST96}}}).
 The number of open
channels is $M=21$ in both (a) and (b) cases.
}
\label{fig8mss}
\end{figure}
\begin{figure}
\caption{Comparison of the spacing distribution $P(s)$ for the $S$ matrix
as defined by Eq.~  
 (\protect{\ref{Smat}})
 [panel (a)] and that for
$S'$ defined via Eq.  
 ~(\protect{\ref{Sma2}})
 [panel (b)]
for long samples $L=800$ and $M=5$. While the latter is well
approximated by the Wigner distribution (a thin solid line in 
panel (b)), the former shows much better agreement with the Berry-Robnik
distribution (a thin solid line in panel (a)) 
corresponding to a superposition of two independent
random ensembles as further discussed in text.
 }
\label{spac3}
\end{figure}
\begin{figure}
\caption{The number variance, $\Sigma^2(d)$ for the reflection sub--matrix $r$
in the deep localization regime of
the scattering: $L=800$ and $M=5$ (a heavy solid line).
 Note a nice agreement
 with the COE prediction (a thin solid line).
 For shorter samples $L=200$ (a heavy dashed line), the number variance
 shows disagreement with the COE behaviour for $d>1.5$.
}
\label{sigma}
\end{figure}

\begin{references}

\bibitem{ALW91} {\sl Mesoscopic Phenomena in Solids}, B.~L.~Altshuler, P.~A.~Lee,
and R.~A.~Webb Eds., North Holland, Amsterdam, 1991.

\bibitem{AMPZ95}E.~Akkermans, G.~Montambaux, J.~L. Pichard and
J.~Zinn--Justin, Eds. Les Houches Session LXI 1994, {\sl Mesoscopic Quantum
Physics}, Elsevier Science B.V., Amsterdam 1995.

\bibitem{Anderson58} P.~W.~Anderson, {\it Phys. Rev.} {\bf
109} 1492 (1958).

\bibitem{abr79} E.~Abrahams, P.~W.~Anderson, D.~C.~Licciardello, and
T.~V.~Ramakrishnan,
     {\it  Phys. \ Rev. \ Lett.} {\bf 42}, 673 (1979).

\bibitem{Ando88} see e.g. {\sl Anderson Localization}, Proceedings of 
the International
Symposium, Tokyo, Japan August 16-18, 1987,  T.~Ando and
H.~Fukuyama, Eds., Springer-Verlag, Berlin, Heidelberg, 1988 and
references quoted in the Introduction.

\bibitem{Mehta}
M. L. Mehta  {\sl Random Matrices, 2 ed.}
 (Academic Press: New York, 1991).

\bibitem{Bena97} C.~W.~J.~Beenakker,
 {\it Rev. Mod. Phys.} {\bf 69}, 731 (1997).

\bibitem{BS88} R.~Bl\"umel and U.~Smilansky,
         {\it Phys. \ Rev. \ Lett. } {\bf 60}, 477 (1988).

\bibitem{Smil90}  U.~Smilansky in {\sl Chaos and Quantum Physics},
Les--Houches Session LII 1989, Ed. M.-J. Giannoni and A. Voros,
North--Holland: Amsterdam 1991

\bibitem{BM94}  H.~U.~Baranger and P.~A.~Mello,
                {\it  Phys. \ Rev. \ Lett. } {\bf 73} 142 (1994).

\bibitem{JPB94} P.~A.~Jalabert, J.-L.~Pichard, and C.~W.~J.~Beenakker,
        {\it Europhys. \ Lett. } {\bf 27}, 255 (1994).

\bibitem{MI95} K. A. Muttalib and M. E. H. Ismail,
 {\sl J. Phys.} {\bf A 28}, L451 (1995).

\bibitem{Stoeckman} H.--J.~St\"ockmann and J. Stein,
{\it Phys. \ Rev. \ Lett.} {\bf 64} 2215 (1990).

\bibitem{FS97} Y.~Fyodorov and H.--J.~Sommers, {\it J. Math. Phys.}
 {\bf 38} 1918 (1997).

\bibitem{wblr97}D.~S.~Wiersma, P.~Bartolini, A.~Lagendijk, and R.~Righini,
                 {\it Nature}, {\bf 390}, 671 (1997).

\bibitem{BG92} F.~Borgonovi and I.~Guarneri,
                {\it J.~Phys. A }{\bf 25}, 3239 (1992).

\bibitem{BG93} F.~Borgonovi and I.~Guarneri,
                 {\it Phys.\ Rev.\ } {\bf E 48}, R2347 (1993).

\bibitem{JP95} P.~A.~Jalabert and J.-L.~Pichard,
                {\it J.~Phys. I France} {\bf 5}, 287 (1995).

\bibitem{Slevin93} K.~Slevin, {\it Electron Transport and
Localisation in Disordered Solids}, Proc. QCTM, Fukui (1993),
 {\it  J. Phys. Soc. Jpn.} {\bf 63} Suppl. A, 170 (1994).

\bibitem{EGST96} P.~Exner, P.~Gawlista, P.~{\v S}eba and
M.~Tater, {\it Ann.  Phys.} {\bf 252} 133 (1996).

\bibitem{AGHH} S.~Albeverio, F.~Gesztesy, R. Hoegh-Krohn, and H~Holden,
{\sl Solvable Models in Quantum Mechanics}, Springer-Verlag,
New York, Berlin, Heidelberg 1988.

\bibitem{EXSE} P.~Exner, P.~{\v S}eba, {\it Phys. Lett} {\bf A222} 1 (1996). 

\bibitem{seba90} P.~\v{S}eba, {\it Phys. Rev. Lett.}  {\bf 64} 1855 (1990).

\bibitem{Seba97a} P.~Exner, P.~\v{S}eba, A.~F.~Sadreev, P.~St\v{r}eda,
and  P.~Feher, {\it Phys. Rev. Lett.} {\bf 80}, 1710 (1998).

\bibitem{lw92} C.~H.~Lewenkopf and H.~A.~Weidenm\"uller,
                 {\it Phys. Rev. Lett.}  {\bf 68}, 3511 (1992).

\bibitem{Dy62}
 F. J. Dyson, {\it J. Math.\ Phys.}  {\bf 3} 140 (1962).

\bibitem{DP} O.~N.~Dorokhov, {\it Pis'ma Zh. Eksp. Teor. Fiz.}
{\bf 36} 259 [ {\it JETP Lett.} {\bf 36} 318 ] (1982).

\bibitem{italo} G. Casati, I. Guarneri, and  G. Maspero,
{\it J. Physique I (France)} {\bf 7}, 729 (1997).

\bibitem{Imry86} Y.~Imry, {\it Europhys. Lett. } {\bf 1} 249 (1986).

\bibitem{Haake} F.~Haake, {\sl Quantum Signatures of Chaos},
    Springer, Berlin, 1991.

\bibitem{fm85} W.~A.~Friedman and P.~A.~Mello,
               {\it  Ann.\ Phys. } {\bf 161} 276 (1985).
               
\bibitem{Bohigas} O.~Bohigas, in: {\sl Chaos and Quantum
Physics}, Les--Houches Session LII 1989, ed.
 M.~J.~Giannoni and A.~Voros, North--Holland, Amsterdam,
1991.

\bibitem{bergmann82} G.~Bergmann, Phys. Rev. {\bf B 25}, 2937 (1982).

\bibitem{MAS88}
P.~A.~Mello, E.~Akkermans and B.~Shapiro,
{\it Phys. \ Rev. \ Lett. } {\bf 61}, 459 (1988).

\bibitem{MS91} P.~A.~Mello and A.~D.~Stone, {\it Phys. Rev. } 
{\bf B 44}, 3559 (1991).

\bibitem{BR84}  M.~V.~Berry and M.~Robnik, {\it J. Phys}
{\bf A 17},  2413 (1984).

\bibitem{PZK97}  M.~Po{\'z}niak, K.~{\.Z}yczkowski, and
M.~Ku{\'s}, {\it J. Phys.} {\bf A 31}, 1059 (1998).
               
\bibitem{GMMB96}  V.~A.~Gopar, M.~Martinez, P.~A.~Mello, and
H.~U.~Baranger,
{\it J. Phys.} {\bf A 29}, 881 (1996).

\bibitem{Z97} K.~{\.Z}yczkowski, {\it  Phys. Rev.}
  {\bf E57}, 2257 (1997).
 
\bibitem{SZZ96} P.~{\v S}eba, K.~\.Zyczkowski and J.~Zakrzewski,
          {\it Phys. Rev.} {\bf E 54} 2438 (1996).

\bibitem{GM98} V.~A.~Gopar and P.~A.~Mello,
         {\it Europhys. Lett.} {\bf 42}, 131 (1998).
 
\bibitem{Mailly92} D.~Mailly and M.~Sanquer,
      {\it J. Phys. I France} {\bf 2} 357 (1992).
      
\bibitem{AHK96} S.~Albeverio, F.~Haake, P.~Kurasov, M.~Ku\'s,
        and P.~{\v S}eba,
        {\it J. Math. Phys.} {\bf 37} 4888 (1996).
        
\bibitem{Stoff95} U.~Stoffregen, J.~Stein, H.-J.~St\"ockmann,
        M.~Ku\'s, and F.~Haake,
        {\it Phys. Rev. Lett.} {\bf 74} 2666 (1995).
        

\end{references}
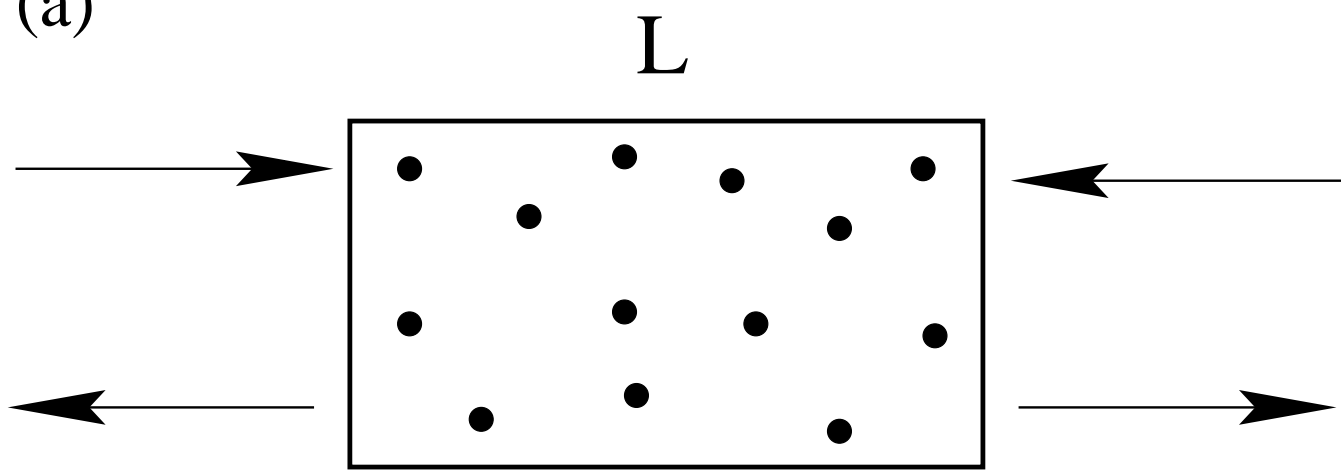
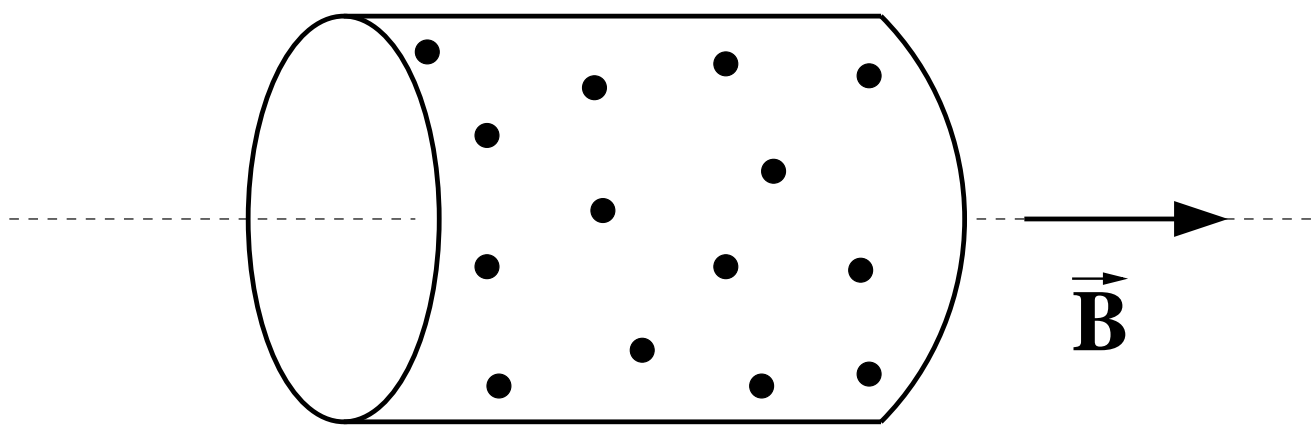
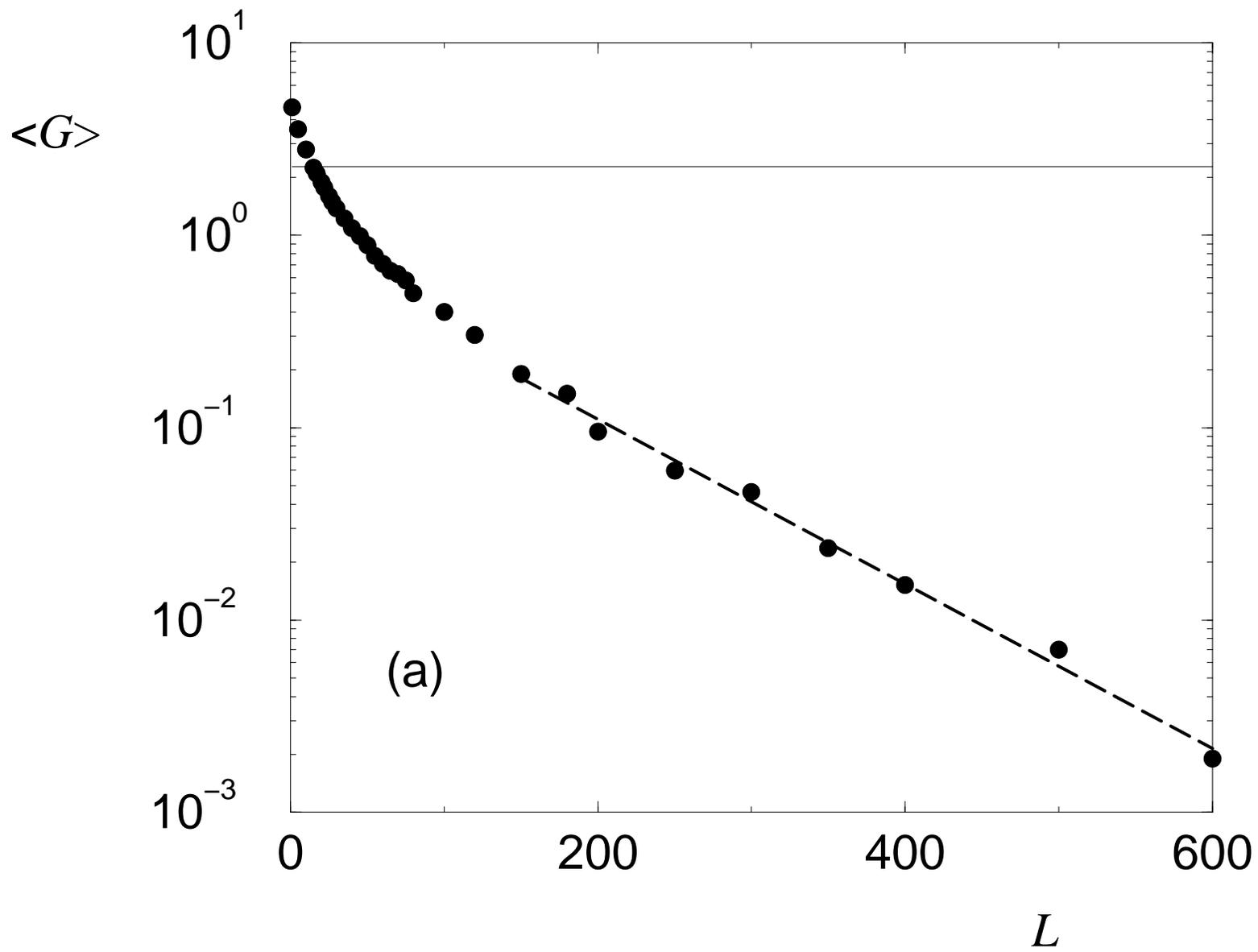
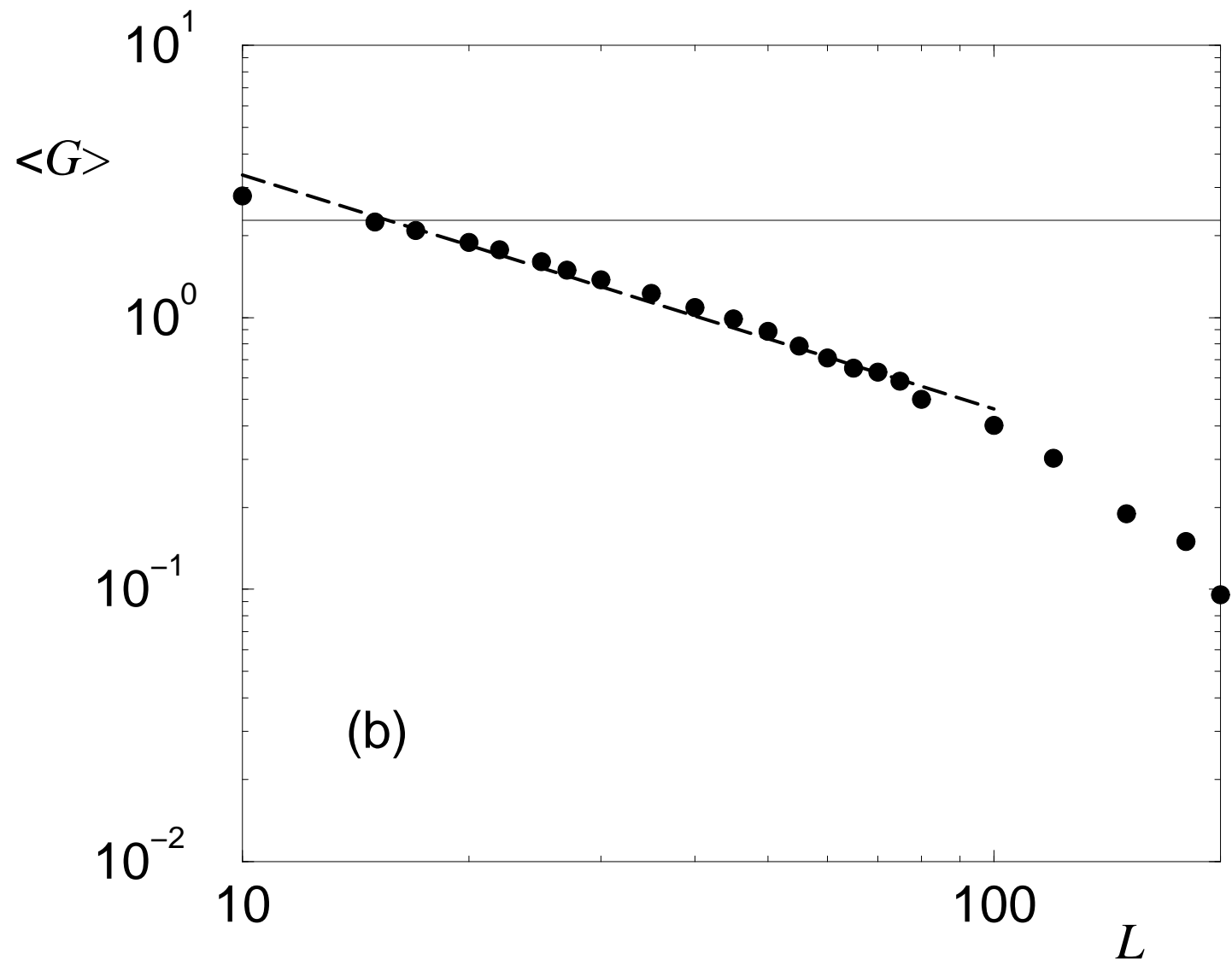
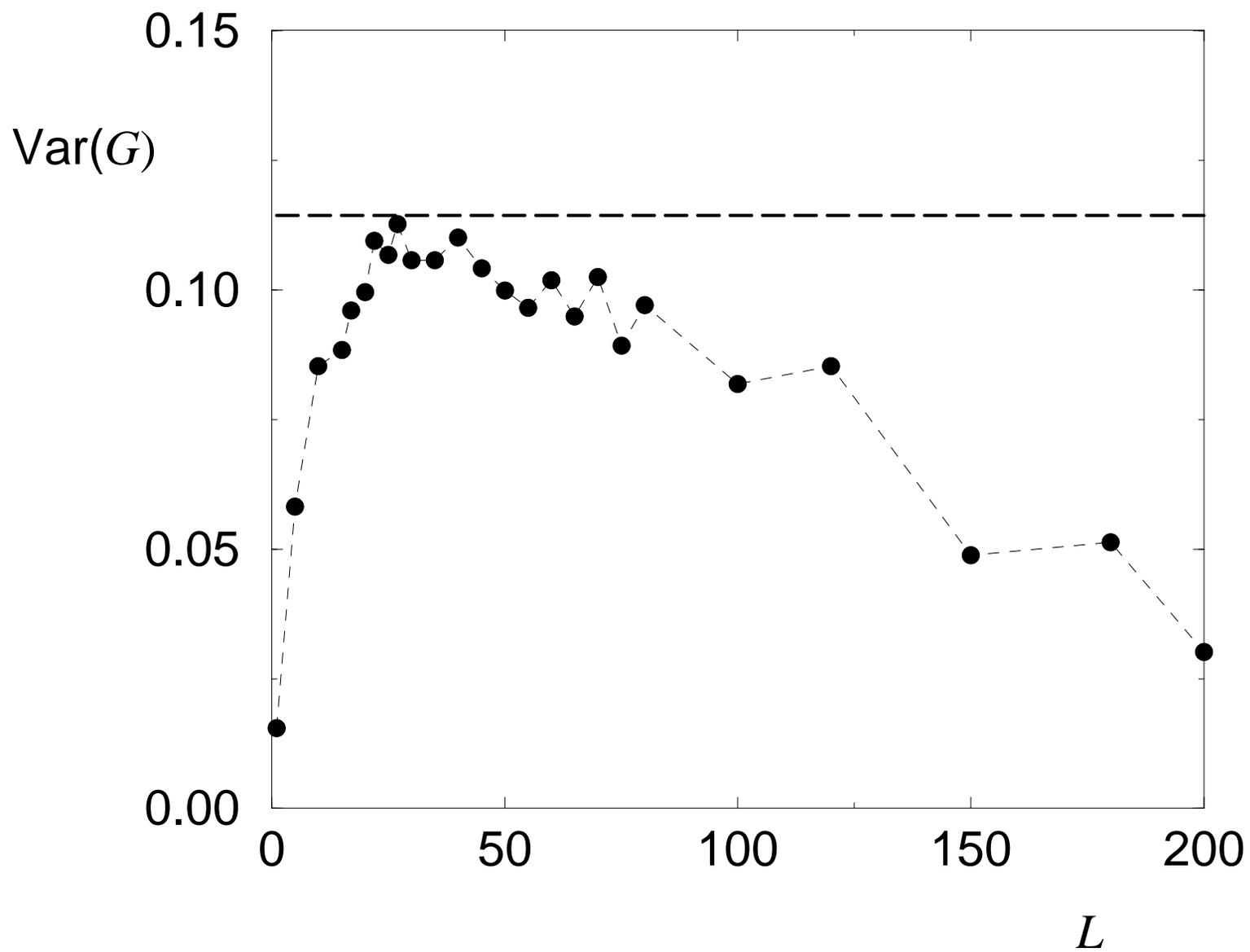
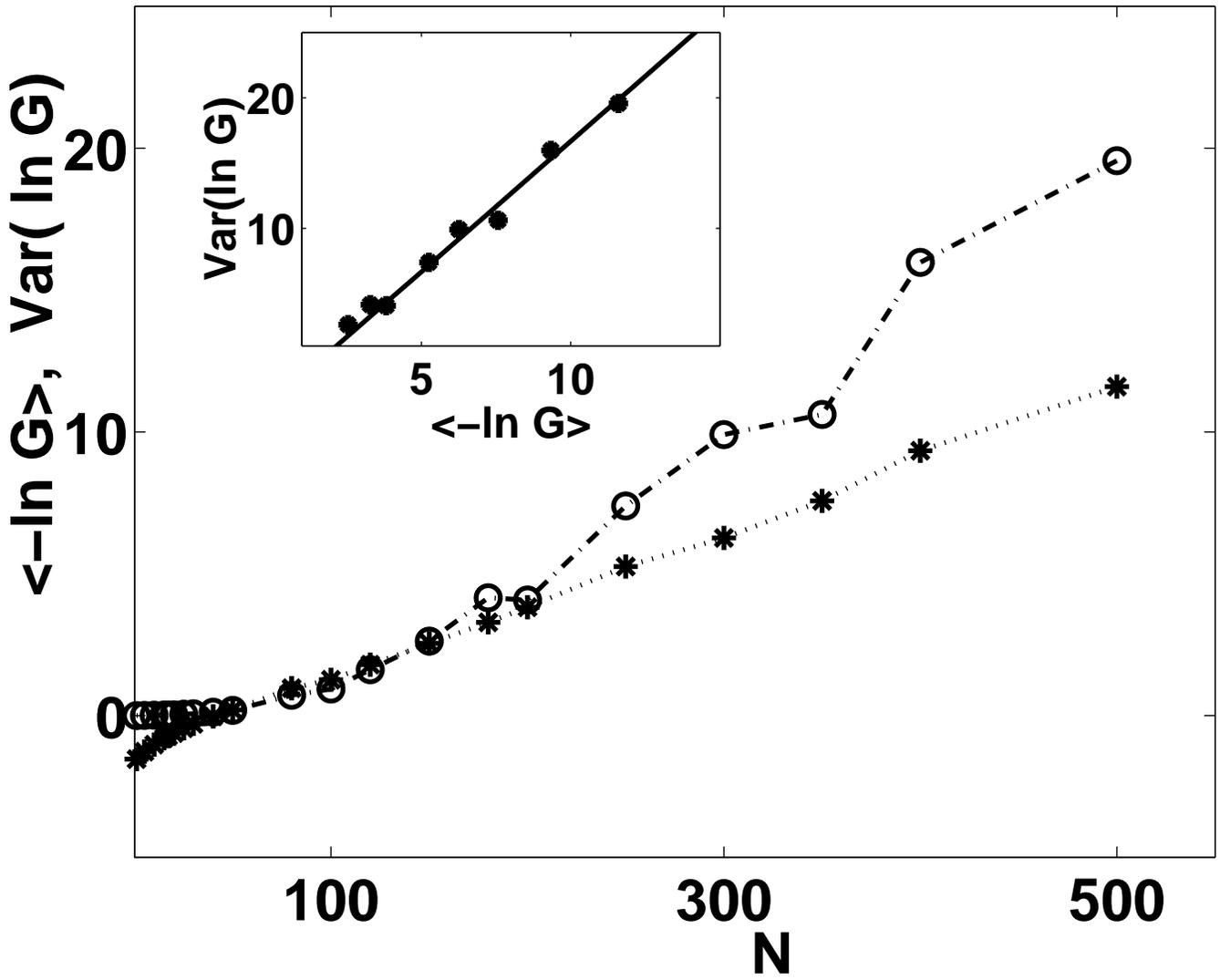
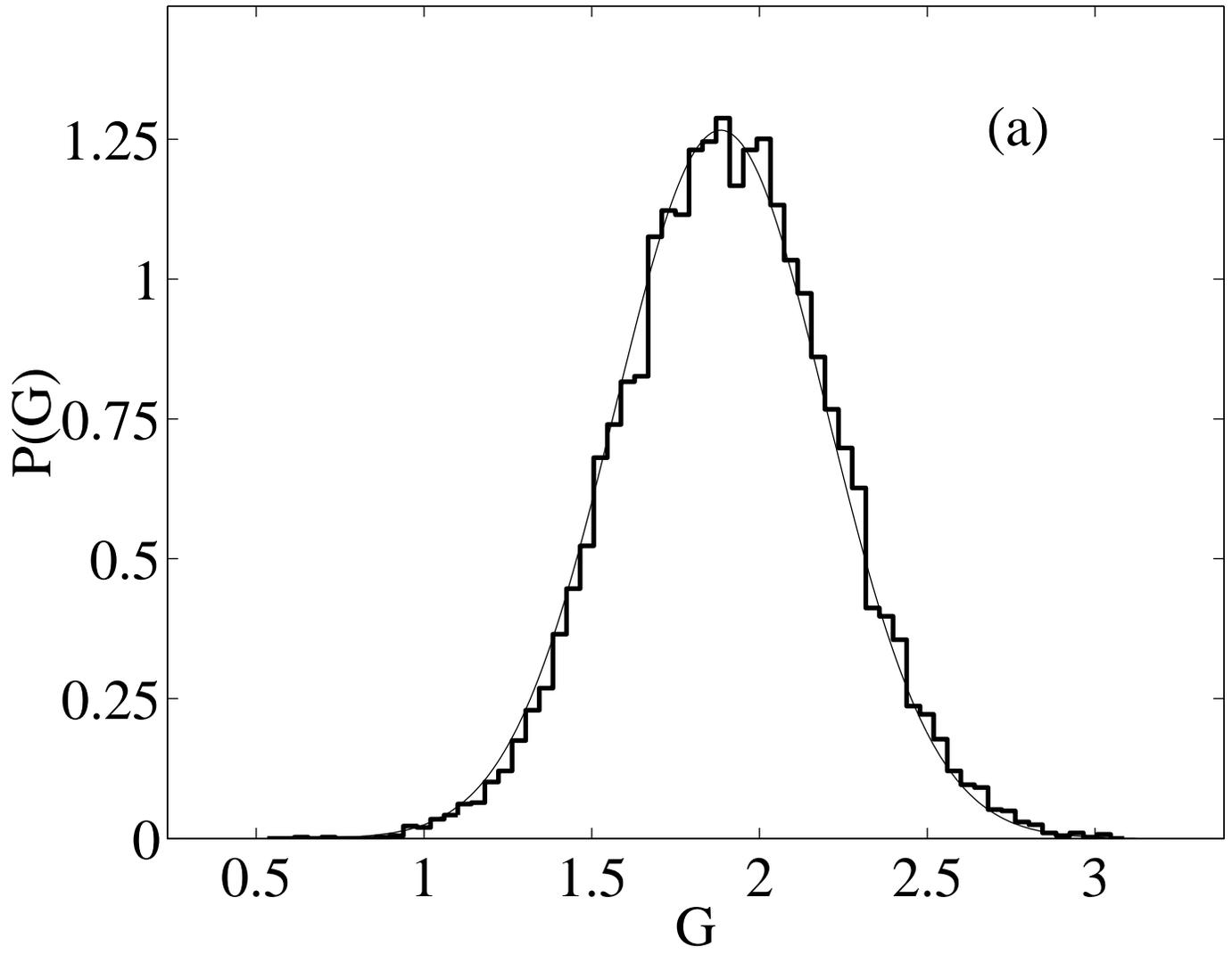
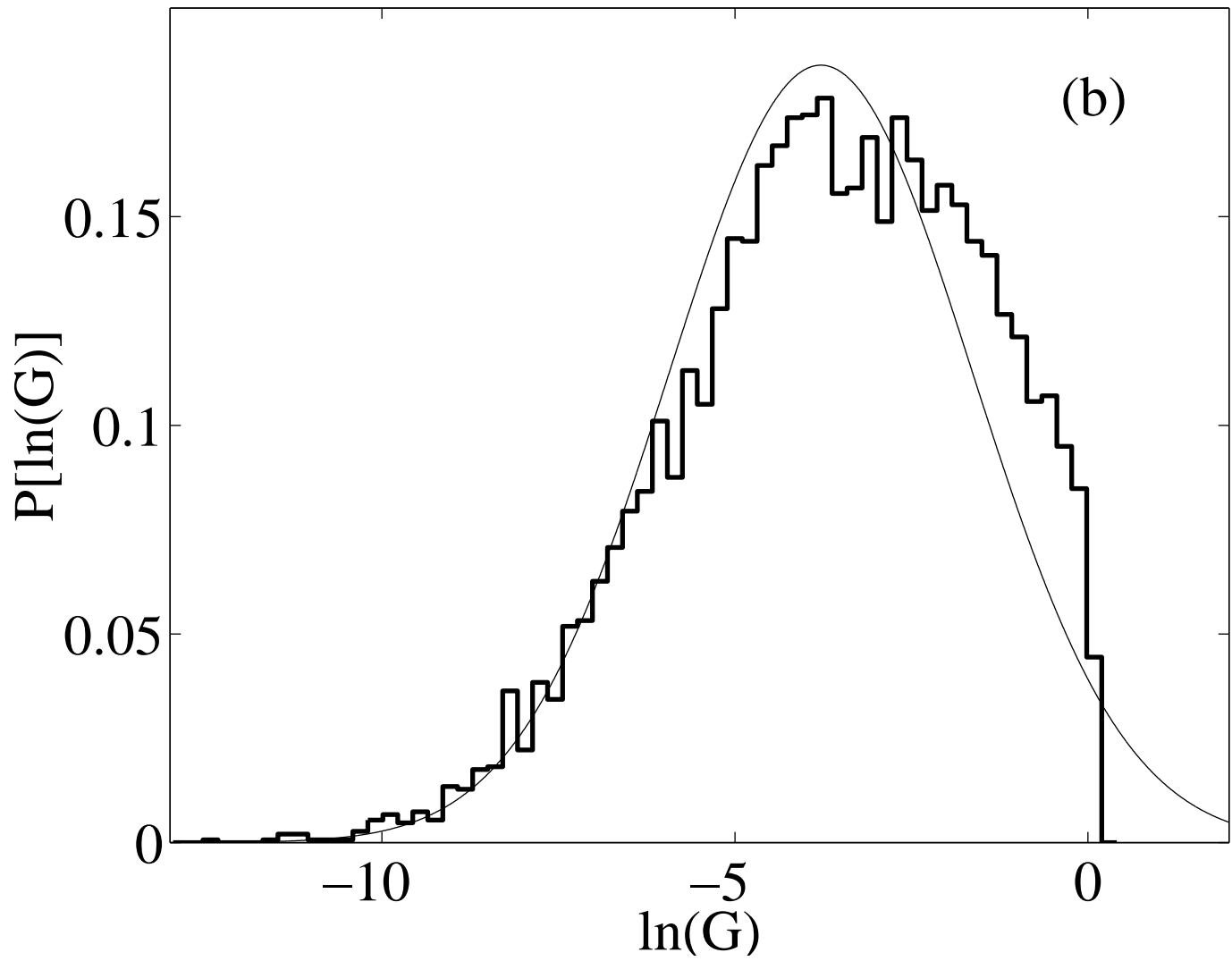
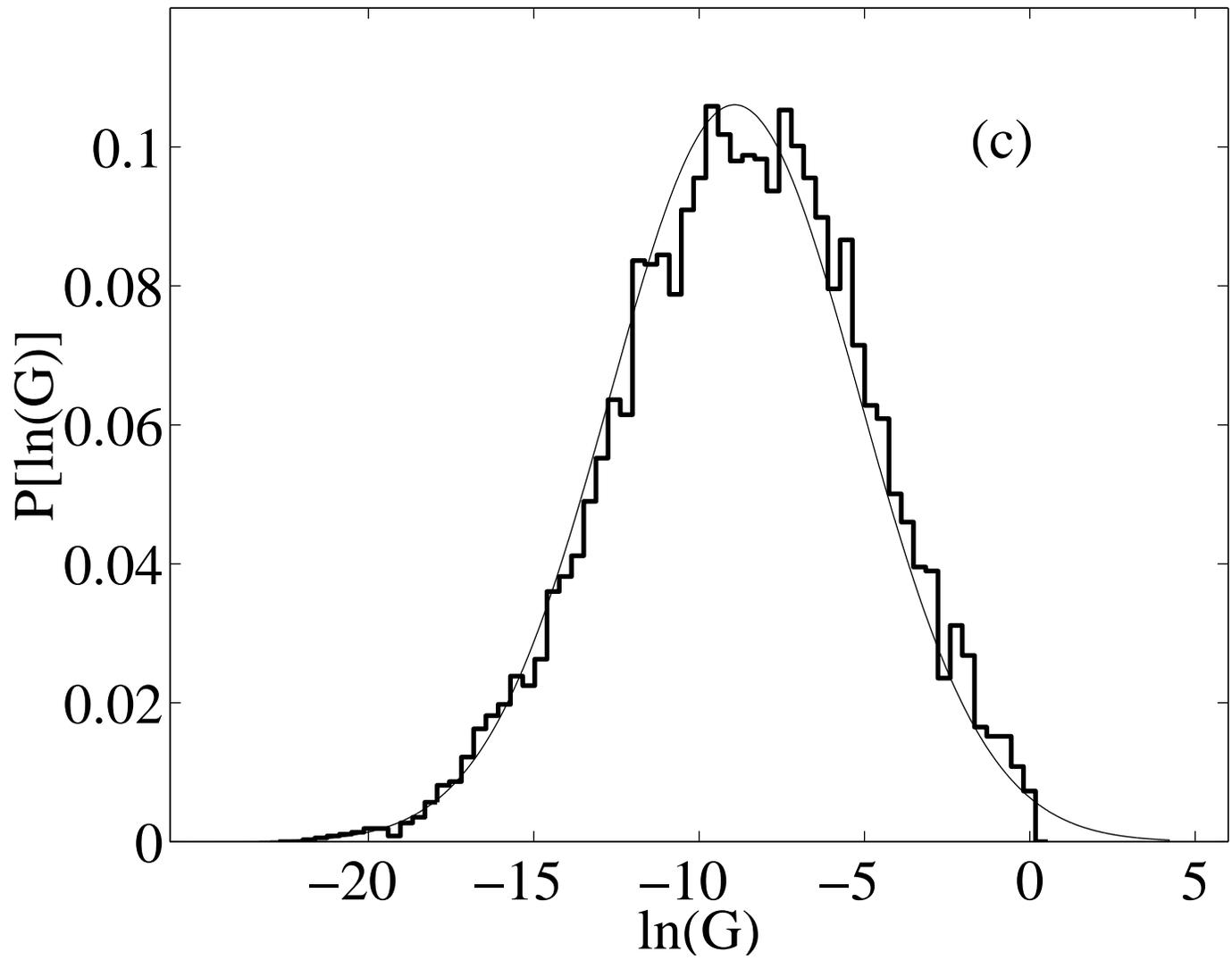
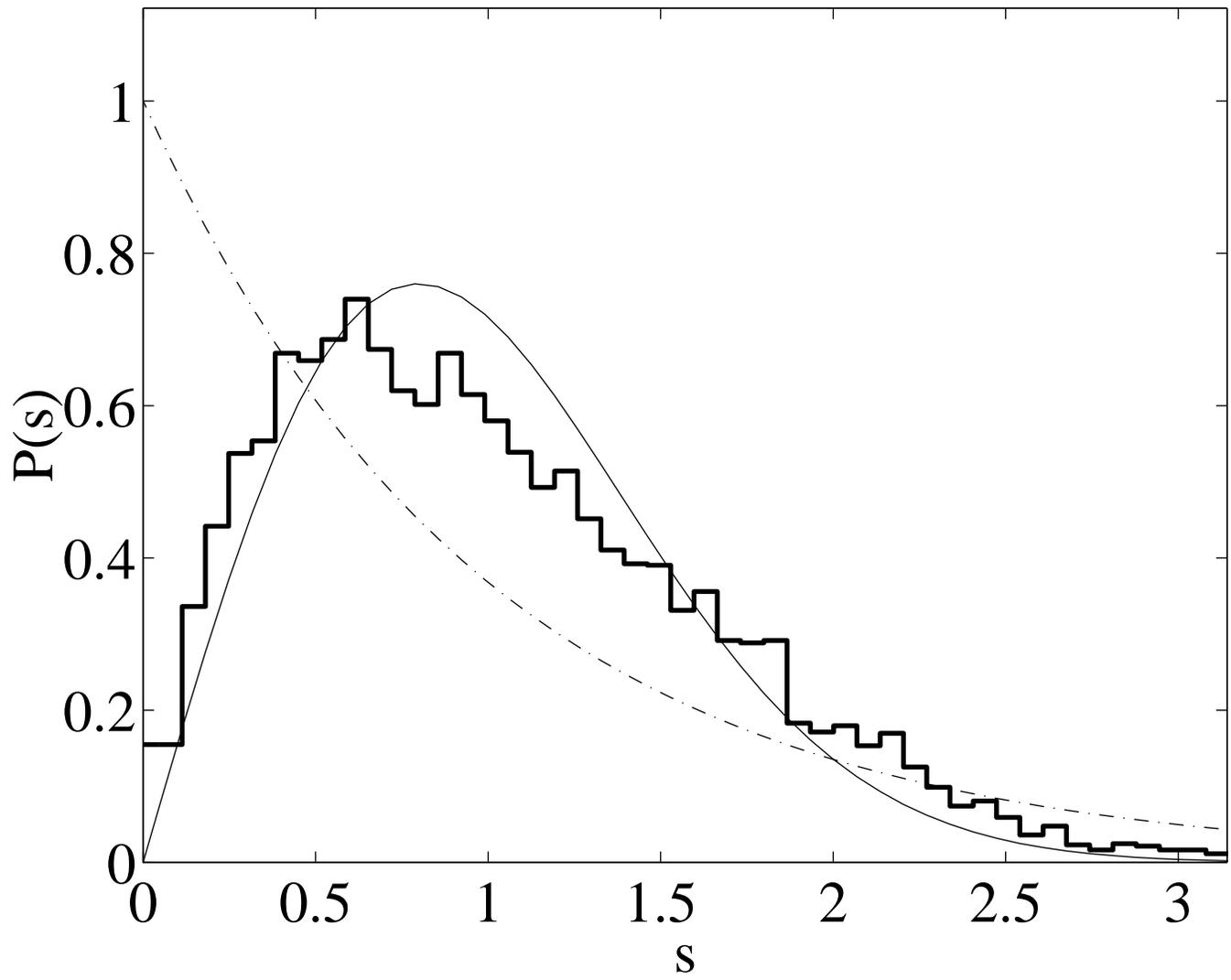
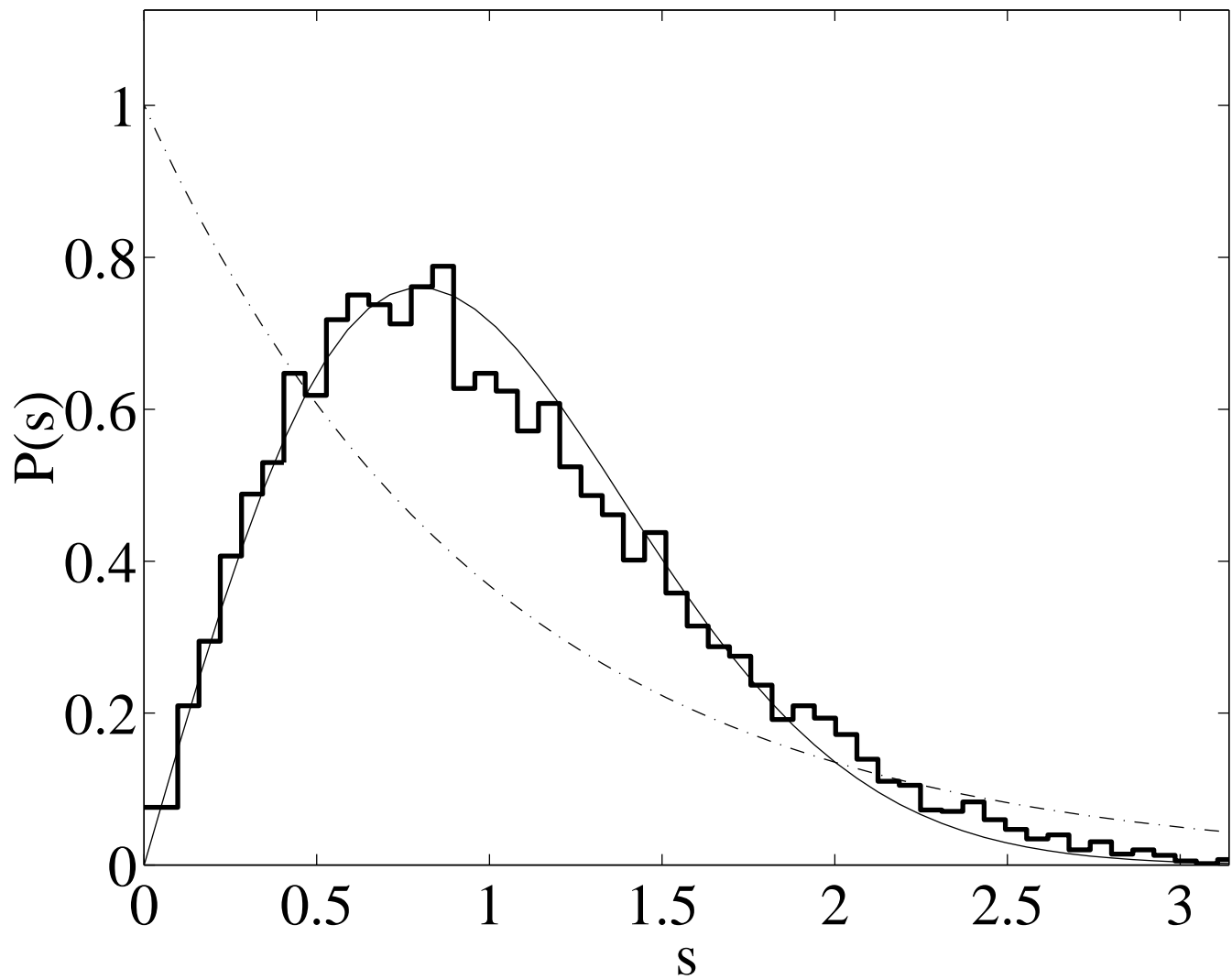
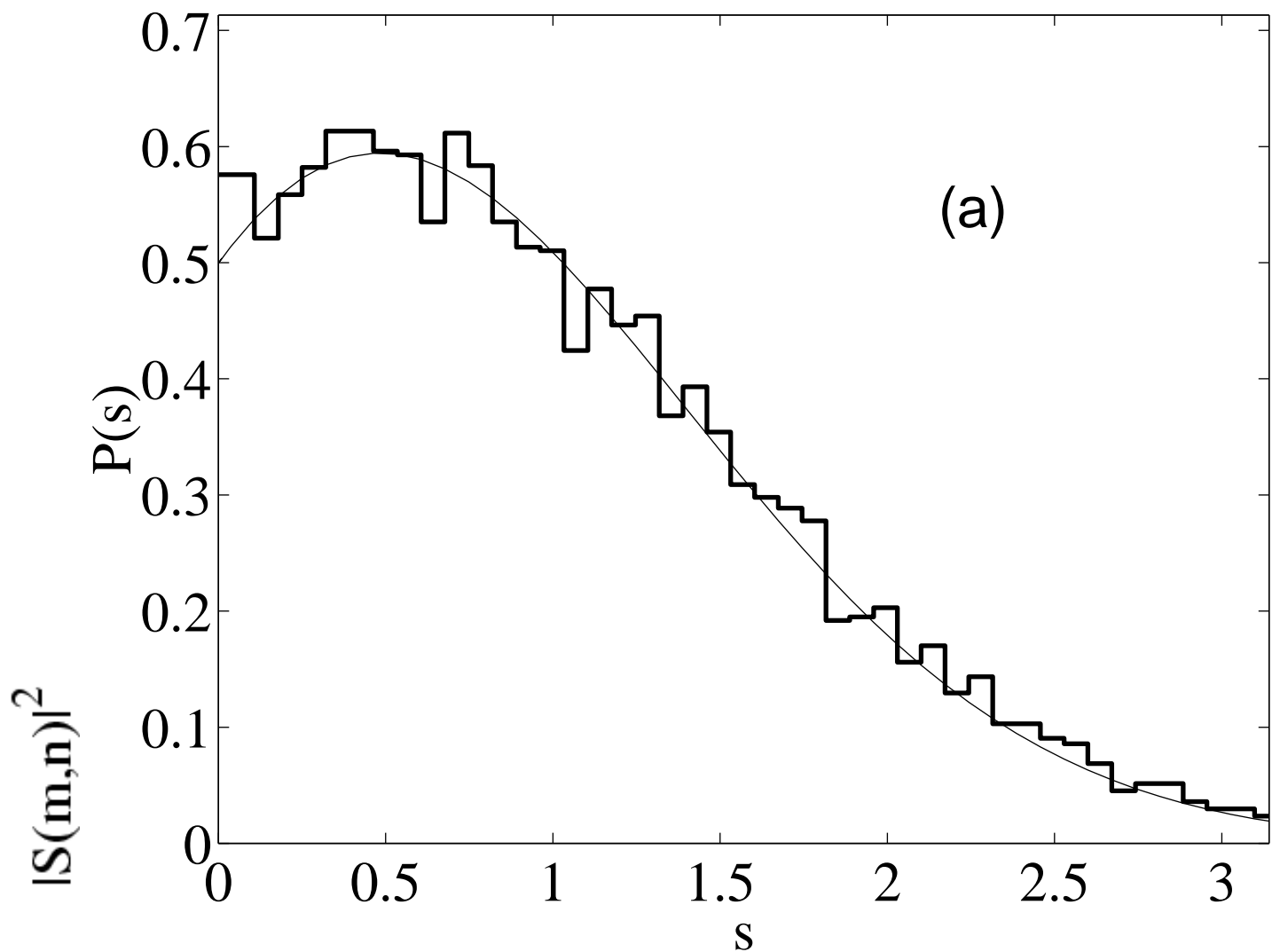
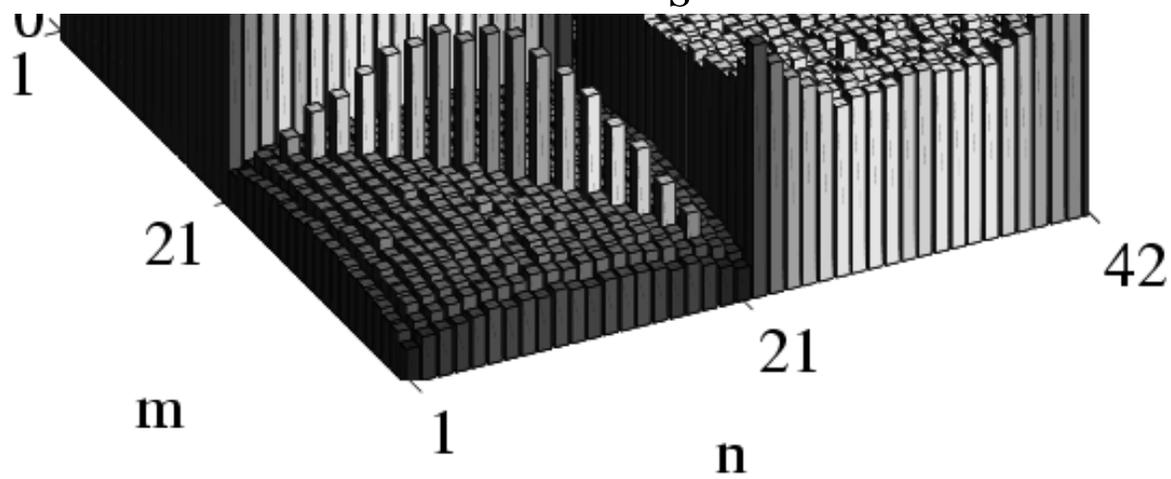
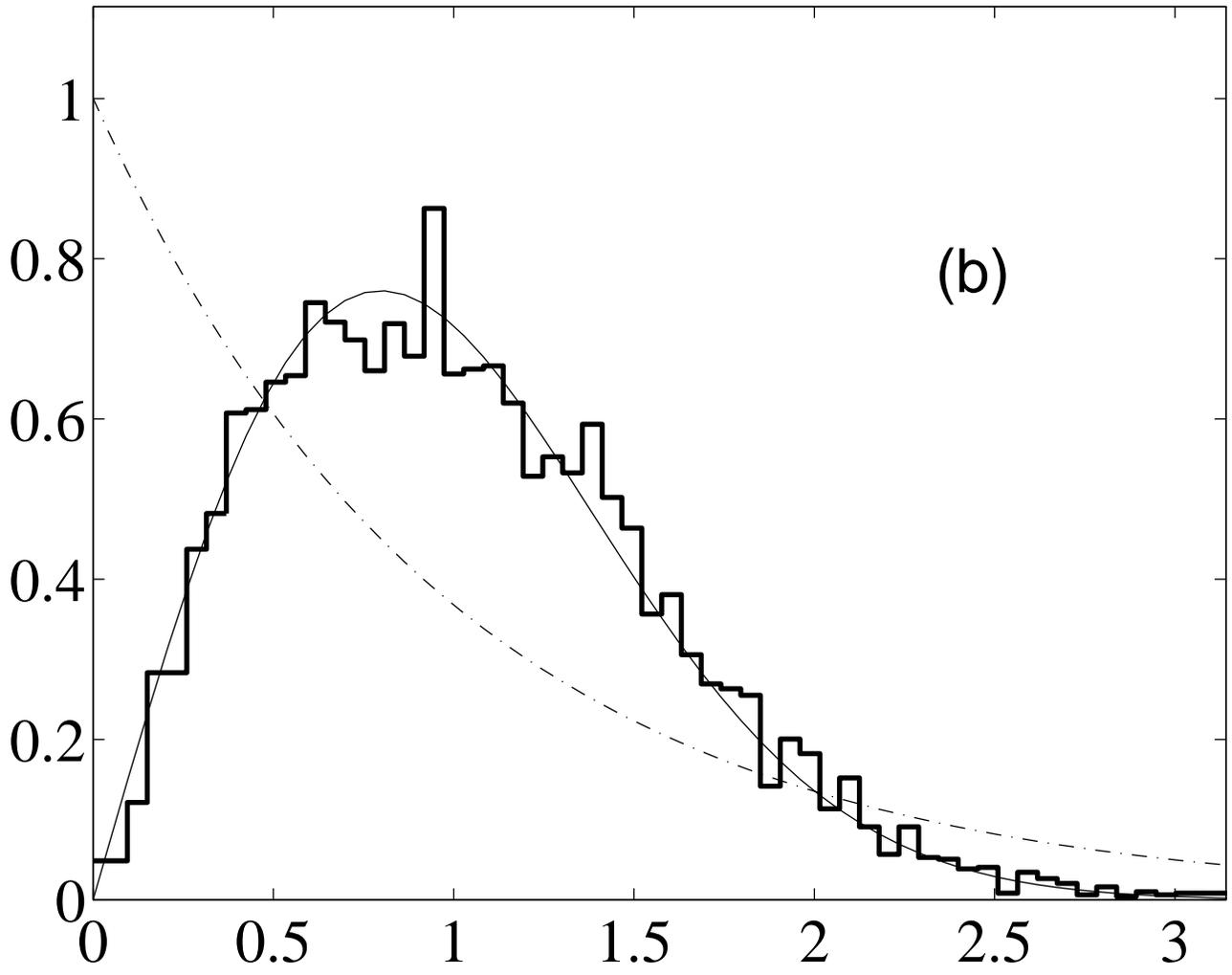
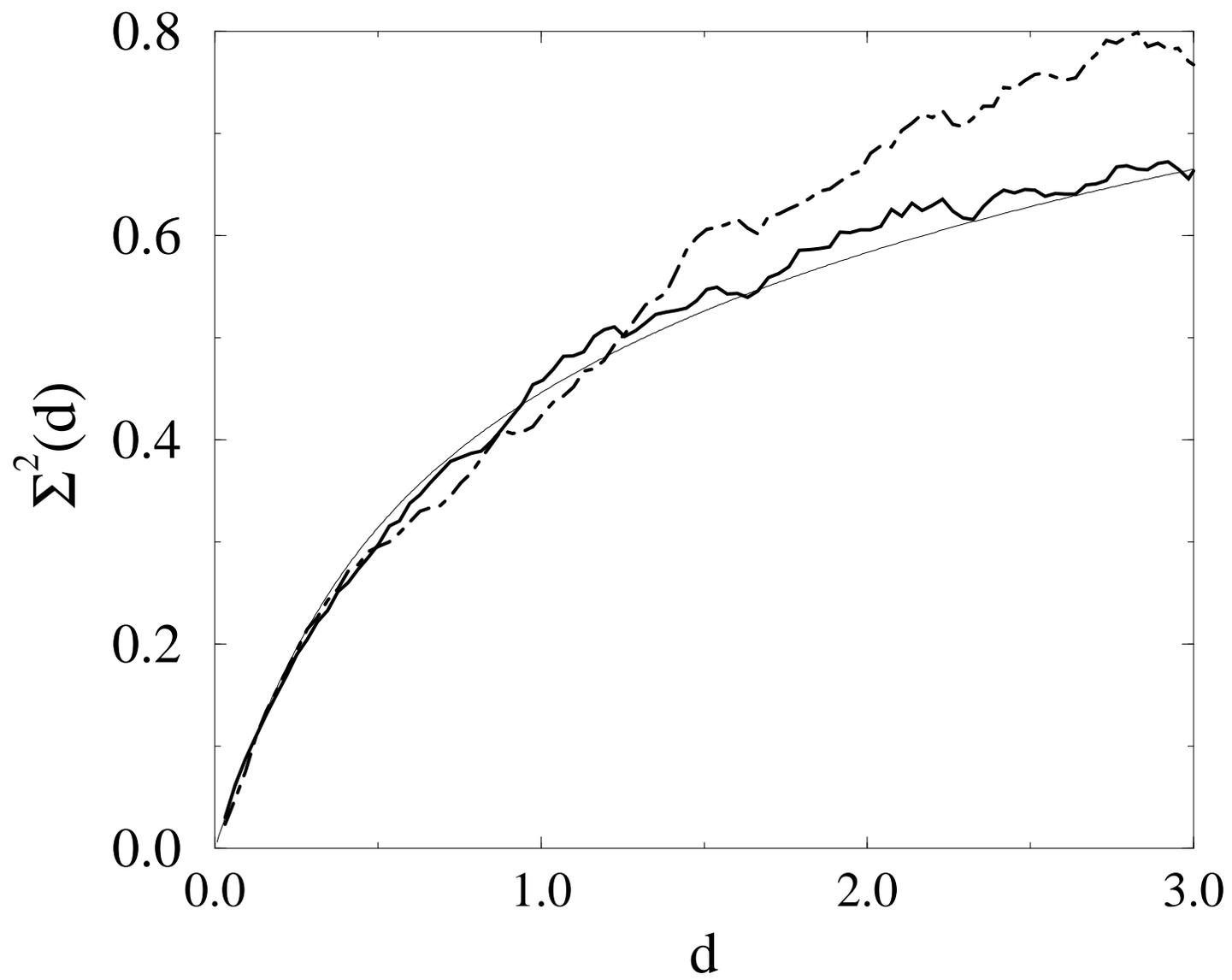
\end{document}